\documentclass[a4paper]{jpconf}
\usepackage{graphicx}
\usepackage{amssymb}
\usepackage{amsmath}
\usepackage{amsfonts}
\usepackage{mathtools}

\usepackage{cite}
\usepackage{ulem}
\usepackage[dvipsnames,svgnames]{xcolor}

\begin{document}
\title{The Vacuum Emission Picture Beyond Paraxial Approximation}

\author{A. Blinne$^{1}$, H. Gies$^{1,2,3}$, F. Karbstein$^{1,2}$, C. Kohlf\"urst$^{1,2}$ and M. Zepf$^{1,4}$}

\address{$^1$Helmholtz Institut Jena, 3, Fr\"obelstieg, Jena, Germany}
\address{$^2$Theoretisch-Physikalisches Institut, Friedrich-Schiller-Universit\"at Jena, 1, Max-Wien-Platz, Jena, Germany}
\address{$^3$Abbe Center of Photonics, 1, Max-Wien-Platz, Jena, Germany}
\address{$^4$Institute of Optics and Quantum Electronics, Friedrich-Schiller-Universit\"at Jena, 1, Max-Wien-Platz, Jena, Germany}

\ead{a.blinne@gsi.de}

\begin{abstract}
Optical signatures of the effective nonlinear couplings among electromagnetic fields in the quantum
vacuum can be conveniently described in terms of stimulated photon emission processes induced by strong
classical, space-time dependent electromagnetic fields. Recent studies have adopted this approach to
study collisions of Gaussian laser pulses in paraxial approximation. The present study extends these
investigations beyond the paraxial approximation by using an efficient numerical solver for the classical
input fields.
This new numerical code allows for a consistent theoretical description of optical signatures of QED
vacuum nonlinearities in generic electromagnetic fields governed by Maxwell’s equations in the vacuum,
such as manifestly non-paraxial laser pulses. Our code is based on a locally constant field approximation
of the Heisenberg-Euler effective Lagrangian. As this approximation is applicable for essentially all optical
high-intensity laser experiments, our code is capable of calculating signal photon emission amplitudes in
completely generic input field configurations, limited only by numerical cost.
\end{abstract}

\section{Introduction}

The vacuum of quantum electrodynamics (QED) is characterized by the omnipresence of quantum fluctuations, effectively providing medium-like properties. 
Charged particle fluctuations can induce effective interactions between electromagnetic fields giving rise to nonclassical effects,
e.g., light-by-light scattering \cite{Euler:1935zz,Karplus:1950zz,Weisskopf}. These vacuum fluctuations supplement Maxwell's linear theory of classical
electrodynamics with effective nonlinear interactions, which invalidate the superposition principle for electromagnetic fields.

These nonlinear corrections become dominant for electromagnetic fields
exceeding the critical photon energy $\omega_{\rm cr} = 2 m_ec^2/\hbar$,
the critical electric field strength $E_{\rm cr}=m_e^2c^3/(e\hbar) \approx 1.3 \times 10^{16}$V/cm or the critical 
magnetic field strength $B_{\rm cr}=E_{\rm cr}/c \approx 4 \times 10^{9}$T; $m_e \approx 511$keV is the electron mass. These critical values are beyond the specifications of all 
currently available optical and x-ray laser systems. Nevertheless, peak field strengths of the order of $E=10^{-3} E_{\rm cr}$ and $B=10^{-3} B_{\rm cr}$ can be reached
at modern laser facilities, such as CILEX \cite{CILEX}, CoReLS \cite{CoReLS}, ELI \cite{ELI}, SG-II \cite{SG-II} and XFEL at DESY \cite{DESY}.
This is insofar relevant, as field strengths of this order are already large enough to potentially allow for the first detection of all-optical signatures of QED vacuum nonlinearities in dedicated experiments.

For the theoretical analysis of all-optical signatures of quantum vacuum nonlinearities the
so-called ``vacuum emission picture'' provides a particularly intuitive framework \cite{Karbstein:2014fva}. 
Based on it, we have devised a highly efficient numerical algorithm allowing for the evaluation of photonic signatures of QED vacuum nonlinearities in generic, experimentally realistic laser fields \cite{Blinne:2018nbd}. 
In our approach, signatures of light-by-light scattering processes are encoded in
signal photons induced in the microscopic interaction of at least three laser background photons. As our algorithm automatically keeps track of all possible combinations regarding
kinematics as well as polarizations of the photons involved in the effective interaction process, it provides direct access to each single characteristics of the emission signal.  

In this article, we study all-optical signatures of quantum vacuum nonlinearities in strong electromagnetic fields
employing the Maxwell solver put forward recently in Refs. \cite{Blinne:2018,Blinne:2018nbd} facilitating first principles studies in electromagnetic field configurations exactly fulfilling Maxwell's equations in vacuum.
It has been found, that in particular in scenarios involving the collision of
multiple laser pulses, the signal photons mainly originate in the small interaction region where the pulses collide. These signal photons are then
detected far away from the interaction region making it easy to analyze their kinematic and polarization properties. 

The article is organized as follows: In Sec.~\ref{sec:formalism}, we briefly review the ``vacuum emission picture'' tailored to the theoretical analysis of optical signatures of quantum vacuum nonlinearities. 
Moreover, we detail our choice of the driving laser fields. An individual laser pulse is chosen to closely resemble the fields of a pulsed paraxial Gaussian beam in the vicinity of the beam focus, but propagated self-consistently according to Maxwell's equations in vacuum. 
In Sec. \ref{sec:results1}, we discuss the effect of self-emission
from a single laser beam. In Sec.~\ref{sec:results2}, we study photonic signatures of QED vacuum nonlinearities in high-intensity laser pulse collisions. 
We conclude this manuscript by summarizing our results in Sec. \ref{sec:summary}.

\section{Theoretical considerations}
\label{sec:formalism}

The present study is based on the vacuum emission picture. In this approach, the laser pulses are treated as background fields, while the signal
photons are treated as quantum fields, see Refs. \cite{Karbstein:2014fva,Gies:2017ezf,Gies:2017ygp} for a more detailed presentation. 
Throughout the article we use the Heaviside-Lorentz system with $c=\hbar=1$ and the metric $g_{\mu \nu} = {\rm diag} (-1,+1,+1,+1)$.

\subsection{Formalism}

At leading order in a loop expansion, the zero-to-single signal photon transition amplitude, to a signal photon state $|\gamma_{p}(\vec{k})\rangle\equiv a^\dag_{\vec{k},p}|0\rangle$ with wave vector $\vec k$ and transverse polarization $p \in \{1,2\}$ is given by
\begin{equation}
 {\cal S}_{(p)}(\vec{k})\equiv\big\langle\gamma_p(\vec{k})\big| \Gamma_\text{int}^{(1)}[A(x)] \big|0\big\rangle.
\end{equation}
Here, $\Gamma_\text{int}^{(1)}[ A(x)]$ is the one-loop Euler-Heisenberg effective action \cite{Heisenberg:1935qt,Schwinger:1951nm} and $A(x)$ is 
the gauge potential of the background electromagnetic field. For laser photon energies $\omega\ll\omega_{\rm cr}$, we can base our considerations on a locally constant field approximation (LCFA) of the Heisenberg-Euler effective action formally derived in constant fields, yielding
\begin{equation}
 S_{(p)}(\vec{k})=\frac{\epsilon_{(p)}^{*\mu}(\vec{k})}{\sqrt{2k^0}}\int{\rm d}^4x\,{\rm e}^{{\rm i}kx}\,j_{\mu}(x)\,\biggr|_{k^0=|\vec{k}|}\,, \label{eq:Sp}
\end{equation}
with the single signal photon current
\begin{equation}
 j_\mu(x)=2\partial^\alpha\frac{\partial{\cal L}_\text{HE}^{1\text{-loop}}(F)}{\partial F^{\alpha\mu}},
\end{equation}
and the signal photon polarization vector $\epsilon_{(p)}^{*\mu}(\vec{k})$. Resorting to spherical coordinates, we express the wave vector of the signal photons $\vec{k}$
and the unit vectors perpendicular to it as
\begin{equation}
 \vec{k} = {\rm k} \hat{\vec{e}}_k= {\rm k}
\left(\begin{array}{c}
  \cos\varphi\sin\vartheta \\
  \sin\varphi\sin\vartheta \\
  \cos\vartheta
 \end{array}\right),
\qquad 
\vec{e}_\perp(\beta)=
\left(\begin{array}{c}
  \cos\varphi\cos\vartheta\cos\beta-\sin\varphi\sin\beta \\
  \sin\varphi\cos\vartheta\cos\beta+\cos\varphi\sin\beta \\
  -\sin\vartheta\cos\beta
 \end{array}\right) . 
\label{eq:epol}
\end{equation}
The latter can be decomposed as
\begin{equation}
 \vec{e}_\perp(\beta)=\vec{e}_1(\vec{k})\cos\beta + \vec{e}_2(\vec{k})\sin\beta ,
 \label{eq:ebeta}
\end{equation}
where we introduced the two basis vectors $\vec{e}_i(\vec{k})$ with $i\in\{1,2\}$.
Equation \eqref{eq:ebeta} can be employed to span the two transversal polarizations of signal photons with wave vector $\vec k$ as
\begin{equation}
 \epsilon^\mu_{(p)}(\vec{k})=\bigl(0,\vec{e}_\perp(\beta_p)\bigr)\,,\quad\text{with}\quad \beta_p=\beta_0+\frac{\pi}{2}(p-1) \, \quad\text{and}\quad \beta_0 \in \mathbb{R}.
\label{eq:polv1&2}
\end{equation}
Finally, upon plugging eq. \eqref{eq:polv1&2} into eq. \eqref{eq:Sp} and limiting ourselves to the leading contribution in a perturbative expansion in the field strengths of the driving laser fields, we can express the signal photon transition amplitude as
\begin{equation}
{\cal S}_{(p)}(\vec{k}) =  \frac{1}{\rm i}\frac{1}{2\pi}\frac{m_e^2}{45}\sqrt{\frac{\alpha}{\pi}\frac{\rm k}{2}}\Bigl(\frac{e}{m_e^2}\Bigr)^3 
 \, \Bigl\{\cos\beta_p\bigl[{\cal I}_{11}(\vec{k})-{\cal I}_{22}(\vec{k})\bigr] 
  +\sin\beta_p\bigl[{\cal I}_{12}(\vec{k})+{\cal I}_{21}(\vec{k})\bigr]\Bigr\} ,
 \label{eq:S1pert}
\end{equation}
where we made use of the definition
\begin{equation}
 {\cal I}_{ij}(\vec{k})=\int{\rm d}^4 x\, {\rm e}^{{\rm i}(\vec{k}\cdot\vec{x}-{\rm k}t)}\,\vec{e}_i\cdot\vec{U}_j\,, \label{eq:Iij}
\end{equation}
with
\begin{align}
 \vec{U}_1=2\vec{E}(\vec{B}^2-\vec{E}^2)-7\vec{B}(\vec{B}\cdot\vec{E}) \,, \qquad
 \vec{U}_2=2\vec{B}(\vec{B}^2-\vec{E}^2)+7\vec{E}(\vec{B}\cdot\vec{E}) \,.
  \label{eqn:Uj}
\end{align}
The differential number of signal photons of polarization $p$ is then given by \cite{Karbstein:2014fva}
\begin{equation}
{\rm d}^3N_{(p)}(\vec{k})=\,\frac{{\rm d}^3k}{(2\pi)^3}\bigl|{\cal S}_{(p)}(\vec{k})\bigr|^2 \,. \label{eq:d3Np_polarcoords}
\end{equation}
Consequently, the total number of signal photons is determined by $N_{\rm tot}=\sum_{p=1}^2 \, \int {\rm d}^3N_{(p)}(\vec{k})$. 

\subsection{Laser fields}
\label{sec:laserfields}

\begin{figure}[t]
\center
\includegraphics[width=8.0cm]{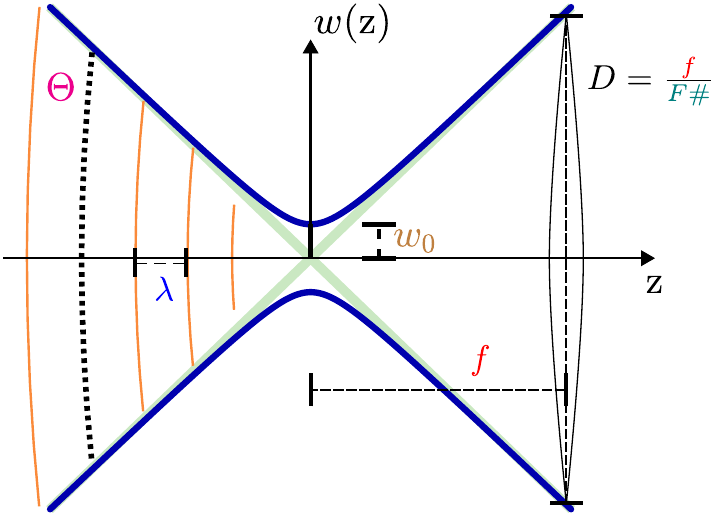} 
\caption{Illustration of the transverse field amplitude profile of a Gaussian laser pulse as a function of the longitudinal coordinate z. The parameter $w_0$ denotes the beam waist, $\lambda$ is the wavelength 
of the pulse, $f$ describes the focal length and $F \#$ its f-number. The parameters $f$ and $F \#$ are related via the the entrance pupil $D$. 
The total angular spread of the laser photons in the far field is given by $\Theta$, while $\theta=\Theta/2$ yields the radial beam divergence. 
Picture adapted from Ref. \cite{Gies:2017ygp}.}
\label{fig:gaussian}
\end{figure}

Our approach is very general in the sense that it allows for the determination of photonic signatures of QED nonlinearities from generic background field configurations. The latter are assumed to be generated by macroscopic ensembles of real propagating photons, whose propagation is governed by Maxwell's equations in vacuum \cite{Blinne:2018nbd}: 
A specific background field configuration is specified by an initial set of data points describing the laser fields at a given time $t_0$. These input data are then self-consistently propagated with a Maxwell solver, ensuring them to exactly solve Maxwell's equations in vacuum. 
Here, we choose the initial data for a given laser pulse to closely resemble analytic expressions modeling the electromagnetic fields of a focused Gaussian laser pulse \cite{Salamin:2006}.
More specifically, in this article we model the driving laser pulses by the spectral pulse model of Ref. \cite{Waters:2017tgl} constructed such as to reproduce the zeroth order paraxial result in the limit of weak focusing; for the technical details, see Ref.~\cite{Blinne:2018nbd}. 
In principle, however, no analytical input is needed. Our numerical code \cite{Blinne:2018nbd} can, for instance, be initialized by the output of standard simulation tools, e.g., a particle-in-cell (PIC) simulation.
This is an important advantage compared to previous approaches, where the paraxial approximation was employed in order to model the high-intensity laser pulses, c.f. Refs. \cite{Karbstein:2014fva, Gies:2017ygp,Tommasini:2010fb,King:2012aw,King:2018wtn, Proceedings} for optical signatures of QED vacuum nonlinearities in the collision of two laser pulses.  

In Fig.~\ref{fig:gaussian}, we recall the fundamentals of laser pulse focusing for a Gaussian beam. 
Of special interest is the total angular spread $\Theta \simeq 2 {\rm arctan} (\frac{\lambda}{\pi w_0})$ of the laser photons in the far field. Here, $\lambda$ is the wavelength, related to the laser photon frequency as $\omega=2\pi/\lambda$. 
The smaller the beam waist size $w_0 = \rho \lambda$, with $\rho > 0$, i.e., the stronger the beam is focused, the larger the angular spread $\Theta$.
At leading order paraxial approximation, the electric peak field strength ${\cal E}_{0}$ of a pulsed linearly polarized beam can be expressed in terms of the laser pulse energy $W$ as ${\cal E}_{0}^2\approx8\sqrt{\frac{2}{\pi}}\frac{W}{\pi w_{0}^2\tau}$ \cite{Karbstein:2017jgh},
where $\tau$ denotes the pulse duration. Note, that leading order paraxial fields are characterized by a single propagation direction (namely the direction of its beam axis) and globally fixed electric and magnetic field vectors. 
The latter are perpendicular to each other and to the laser's beam axis.
In this particular case, all nontrivial features of the laser pulse, such as focusing effects, are encoded in a single field profile; cf. Fig. \ref{fig:gaussian} and, e.g., Ref.~\cite{Karbstein:2014fva} for the analytic expressions. 

In this article, we aim at a comparison of the results obtained by means of our numeric Maxwell solver \cite{Blinne:2018nbd} with the corresponding outcomes of established methods. To this end, we focus on the following two scenarios: 
signal photon self-emission from a single focused high-intensity laser pulse in Sec. \ref{sec:results1}, and signal photon emission in the collision of two focused laser beams in Sec. \ref{sec:results2}. 
The former scenario was previously studied in Ref.~\cite{PhysRevLett.107.073602}, and the latter in Ref.~\cite{Gies:2017ygp}.
For completeness, note that leading order paraxial beams do not exhibit a self-emission phenomenon, due to the fact that both scalar invariants, $\vec{B}\cdot\vec{E}$ and $\vec{B}^2-\vec{E}^2$, vanish for such fields; for further details, see Ref.~\cite{Gies:2017ygp}. 

\section{Single-beam scenario}
\label{sec:results1}

In this section, we study the effect of signal-photon self-emission from a single focused high-intensity laser pulse. This phenomenon was proposed as a prospective signature of QED vacuum nonlinearity in Ref.~\cite{PhysRevLett.107.073602}, focusing in particular on the number of emitted signal photons from a laser field focused with a large angular aperture \cite{Richards358}. 
The calculations in Ref.~\cite{PhysRevLett.107.073602} are based on vector diffraction theory; cf. Ref.~\cite{Richards358} for the derivation of the electric and magnetic fields in the vicinity of the beam focus. 

\begin{figure}[t]
\center
\includegraphics[width=10.0cm]{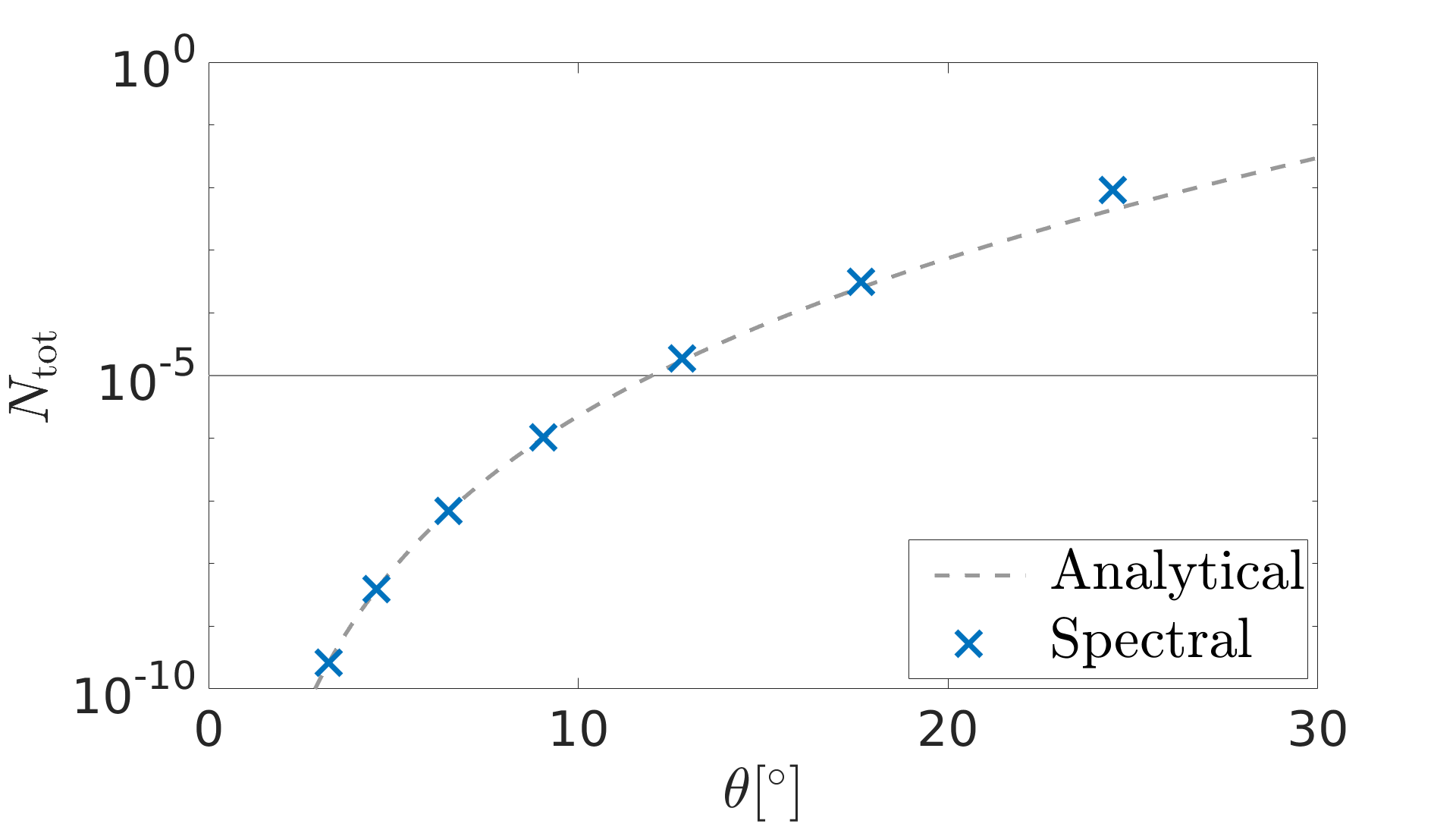} 
\caption{Total number of signal photons $N_{\rm tot}$ emitted by a single Gaussian beam ($\lambda=800$nm, $W=25$J, $\tau=25$fs) 
as a function of the radial beam divergence $\theta$ (blue markers). For comparison, we depict the corresponding analytical scaling behavior~\eqref{eq:scaling} (grey line). The proportionality constant is fixed by demanding the curve to go through the numerical result for the smallest available divergence $\theta$.}
\label{fig:monden}
\end{figure}

In Fig.~\ref{fig:monden},  we plot the total number of signal photons $N_{\rm tot}$ as a function of the radial beam divergence $\theta$ for a high-intensity laser pulse of wavelength $\lambda = 800$nm, pulse duration $\tau = 25$fs
and total pulse energy $W=25$J. The crosses depict our results obtained by our numerical solver using the spectral pulse model \cite{Waters:2017tgl}. As detailed in Sec.~\ref{sec:laserfields}, the far-field divergence of a laser beam is fully determined by the focusing parameters of the pulse as $\theta=\Theta/2\simeq{\rm arctan} (\frac{\lambda}{\pi w_0})$.
Hence, the stronger the pulse is focused, i.e. the smaller the beam waist $w_0$, the larger the number of signal photons. In the limit of unfocused beams, i.e., for $\theta\to 0$, we recover the case of a single plane wave, featuring perpendicular electric and magnetic fields of the same amplitude.
As there is no self-emission from a single plane wave, $N_{\rm tot}$ vanishes for small $\theta$. However, as soon as the pulse
is focused, it can no longer be described by orthogonal electric and magnetic fields of the same amplitude, when demanding it to exactly fulfill Maxwell's equations in vacuum. 
Therefore, effective self-interactions of the laser field are induced stimulating the onset of signal photon self-emission.
The scaling of  the total signal photon number with the radial beam divergence can be estimated from the analytical result obtained from first-order paraxial fields \cite{Blinne:2018nbd,KarbsteinSundqvist},
\begin{equation}
  N_{\rm tot} \sim \left(\frac{\lambda}{w_0}\right)^8 \sim \tan^8 \theta.
  \label{eq:scaling}
\end{equation}
Fixing the proportionality factor of this scaling using the numerical result for the smallest available divergence $\theta$, the analytical scaling curve is shown as a dashed line in Fig.~\ref{fig:monden}; this estimate also matches with the results obtained in Ref.~\cite{PhysRevLett.107.073602}. While the small-angle behavior agrees well with the full numerical result, the deviations occur for increasing beam divergence. Here, the scaling predicted by the first-order paraxial fields is no longer sufficient and underestimates the true behavior obtained by our numerical solver. 
The stronger a laser pulse is focused, the larger the spread of the wave vectors of the laser photons, and thus the stronger the macroscopic electric and magnetic field components of the laser field which are not orthogonal to its beam axis.
Reliable quantitative estimates for self-emission from strongly focused pulses hence requires a self-consistent description of the laser pulse. 

\section{Two-beam scenario}
\label{sec:results2}

\begin{figure}[t]
\center
\includegraphics[width=7.8cm,trim=20 0 90 40, clip]{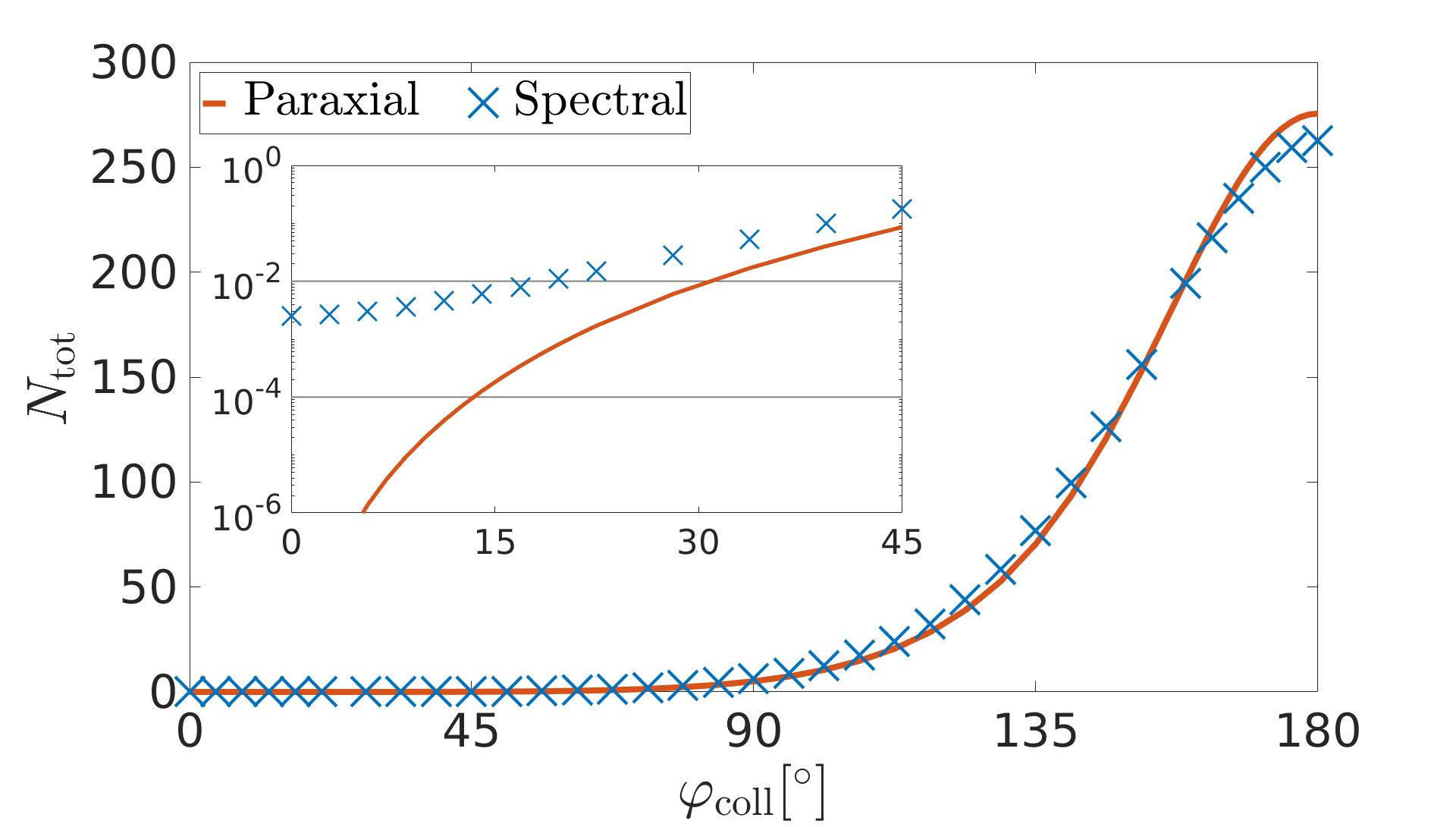} 
\includegraphics[width=7.8cm,trim=20 0 90 40, clip]{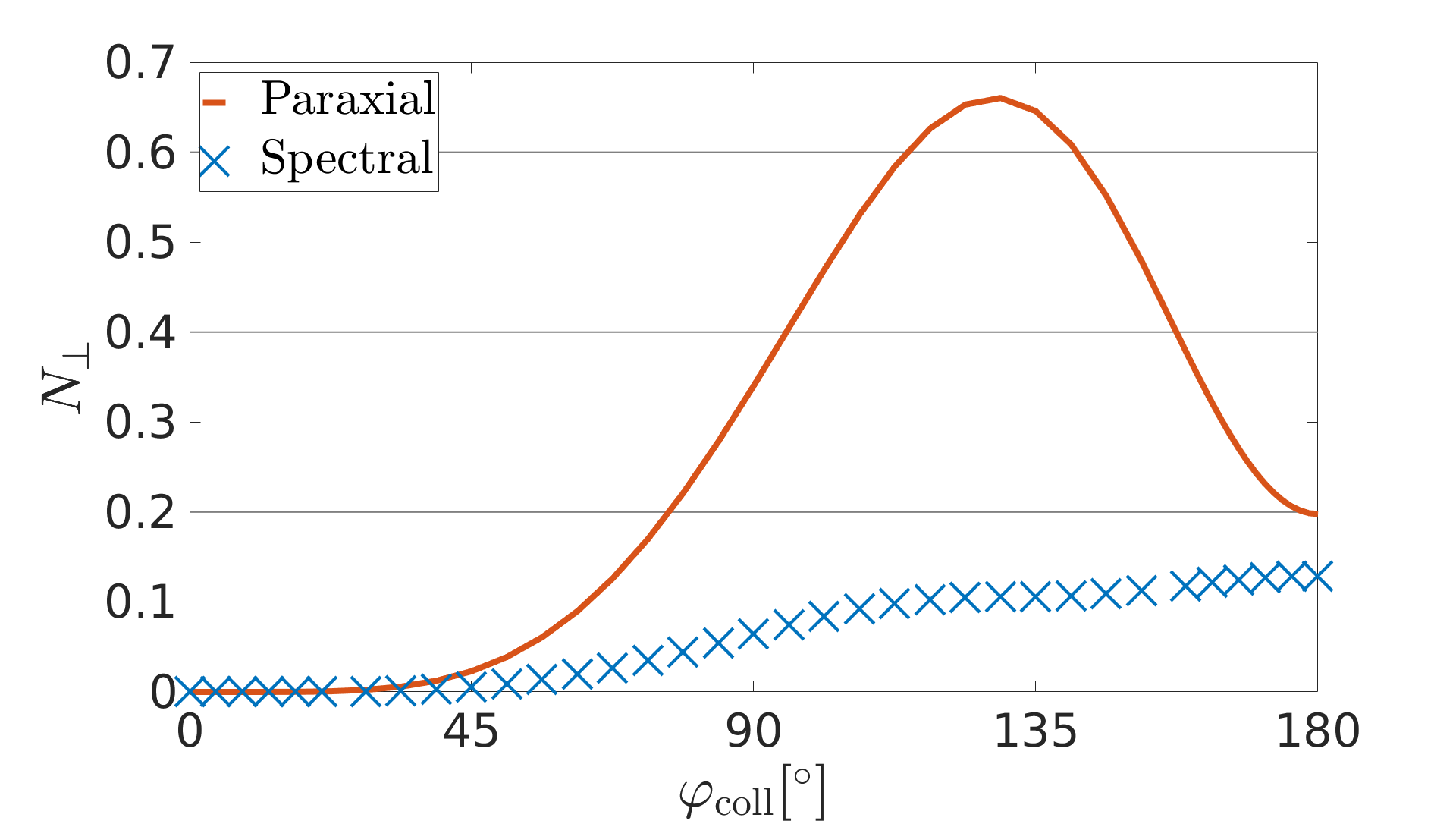} 
\caption{Total number of signal photons $N_{\rm tot}$ (left) attainable for two identical high-intensity laser pulses
($w_0=\lambda=800$nm, $W=25$J, $\tau=25$fs) colliding under 
an angle $\varphi_{\rm coll}$. The inset highlights the substantial deviations of the paraxial and spectral results for small collision angles. 
The differences between the two approaches are particularly pronounced in the results for the number of signal photons polarized perpendicularly to the driving laser pulses in the beam foci $N_\perp$ (right).}
\label{fig:phi}
\end{figure}

Next, we focus on the collision of two identical laser pulses being polarized perpendicular to the collision plane in the beam focus; the beams are assumed to be perfectly synchronized and are focused exactly to the same spot. 
More specifically, we compare results obtained with our numerical code, 
propagating the driving laser fields self-consistently according to Maxwell's equations in vacuum, and using the leading-order paraxial approximation to model the driving laser fields. 
Each of the laser pulses is assumed to have a pulse energy of $W=25$J, a pulse duration of $\tau=25$fs and a wavelength of $\lambda=800$nm. Moreover, both pulses are focused to the diffraction limit, such that $w_0=\lambda$.
Our findings are displayed as a function of the collision angle $\varphi_{\rm coll}$ in Fig. \ref{fig:phi}. Here, we study both the total number of attainable signal photons $N_{\rm tot}$ as well as the number of signal photons 
polarized perpendicularly to the incident laser beams $N_\perp$. 
A vanishing collision angle $\varphi_{\rm coll}=0$ corresponds to two co-propagating beams, while $\varphi_{\rm coll}=180^\circ$ amounts to a head-on collision.

The clear deviations in the result for $N_{\rm tot}$ for small collision angles can be explained by the fact that the zeroth order paraxial approximation does not at all account for the effect of signal photon self-emission.
Nevertheless, for larger values of $\varphi_{\rm coll}$ both approaches yield compatible results for $N_{\rm tot}$, thereby substantiating previous predictions based on the paraxial approximation \cite{Gies:2017ygp}.
Substantial differences occur for the number of perpendicularly polarized signal photons $N_\perp$: a calculation employing
paraxial fields predicts a spurious maximum at $\varphi_{\rm coll} \sim 130^\circ$ rather than a smooth increase in $N_{\perp}$ as predicted by the field configuration self-consistently propagated with our numerical code. 
These deviations can be explained as follows: The quantity $N_{\perp}$ accounts for signal photons being polarized perpendicular to the polarization vector $\vec e_E$ of the incident laser beams in the beam focus. 
Hence, the number of perpendicularly polarized signal photons is given by $N_\perp= \int {\rm d}^3N_{(p)}(\vec{k}) \big \vert_{\beta_p(\varphi, \vartheta)}$ 
with the direction dependent angle $\beta_p(\varphi, \vartheta)$ fixed such that $\vec e_\perp(\beta_p) \cdot \vec e_E=0$ \cite{Karbstein:2014fva, Gies:2017ygp, Blinne:2018nbd}.
Within the paraxial
approximation the polarization vectors of the incident laser pulses are globally fixed. Hence, all laser photons are formally described by a single polarization vector, which -- in our case -- is pointing perpendicular to the collision plane. 
The situation is completely different for the case of laser pulses fulfilling Maxwell's equations in vacuum exactly. Away from the laser focus 
the polarization vectors of the laser photons naturally feature polarization components perpendicular to the polarization vector in the focus. 
Consequently, no global polarization vector exists and differences in the result for $N_\perp$, based on a fixed definition of the polarization direction, are not too surprising. However, these observations clearly underpin the necessity of a careful and critical reassessment of predictions based on paraxial fields before their actual promotion as prospective signatures of QED vacuum nonlinearities; 
for more details we refer to Ref.~\cite{Blinne:2018nbd}. 

\section{Summary}
\label{sec:summary}

In this article, we have exemplified the great potential of our new numerical tool \cite{Blinne:2018nbd} tailored to study all optical signatures of quantum vacuum nonlinearities by applying it to two specific scenarios previously discussed in Refs.~\cite{PhysRevLett.107.073602,Gies:2017ygp}. 
As our code propagates any given initial field configuration self-consistently according to Maxwell's equations in vacuum, it goes substantially beyond many previous studies, modeling the driving laser fields by approximate solutions of Maxwell's equations. 
We have demonstrated that the deviations between the results based on exact and approximate solutions of Maxwell's equations in vacuum to model the driving laser fields can be unexpectedly large for specific scenarios. 
Clear deviations are in particular encountered in the study of critically polarization sensitive quantities, such as the attainable numbers of signal photons scattered into a perpendicularly polarized mode. 

To sum up, the advent of petawatt-class high-intensity laser systems has brought QED vacuum nonlinearities closer to detection than ever before.
Especially in combination with polarization- and direction sensitive single-photon detection schemes, measuring optical signatures of QED vacuum nonlinearities seems to be 
within reach at modern high-intensity laser facilities. This would constitute the first discovery of QED vacuum nonlinearities  in macroscopic electromagnetic fields in a well-controlled laboratory experiment for the first time.
As precision experiments require accurate quantitative theoretical predictions of the effects to be searched for,
we expect our new numerical code to play a major role in this endeavor.

\section*{Acknowledgments}

We are grateful to Nico Seegert and Andr\'{e} Sternbeck for many helpful discussions and
support during the development phase of the numerical algorithm.  The
work of C.K.~is funded by the Helmholtz Association through the
Helmholtz Postdoc Programme (PD-316). We acknowledge support by the
BMBF under grant No. 05P15SJFAA (FAIR-APPA-SPARC).

Computations were performed on the ``Supermicro Server 1028TR-TF'' in Jena, which
was funded by the Helmholtz Postdoc Programme (PD-316).


\section*{References}
\begin{thebibliography}{99}        
	  
    \bibitem{Euler:1935zz} 
      H.~Euler and B.~Kockel,
      Naturwiss.\  {\bf 23}, 246 (1935).
      
\bibitem{Karplus:1950zz} 
  R.~Karplus and M.~Neuman,
  Phys.\ Rev.\  {\bf 83}, 776 (1951).
      
    \bibitem{Weisskopf}
      V.~Weisskopf, 
      Kong.\ Dans.\ Vid.\ Selsk., Mat.-fys.\ Medd.\ {\bf XIV}, 6 (1936).  
     
    \bibitem{CILEX}
      CILEX, http://cilexsaclay.fr/ .
      
    \bibitem{CoReLS}
      CoReLS, http://corels.ibs.re.kr/ .
      
    \bibitem{ELI}
      ELI, https://eli-laser.eu/ .

    \bibitem{SG-II}
    X.~Xie, J.~Zhu, Q.~Yang, J.~Kang, H.~Zhu, M.~Sun and A. Guo,
    CLEO Technical Digest, paper SM1M.7 (2016).
    
    \bibitem{DESY}
      DESY, https://www.desy.de .    
      
    \bibitem{Karbstein:2014fva}
      F.~Karbstein and R.~Shaisultanov,
      Phys.\ Rev.\ D {\bf 91} (2015) no.11,  113002
      [arXiv:1412.6050 [hep-ph]].
      
    \bibitem{Blinne:2018nbd} 
      A.~Blinne, H.~Gies, F.~Karbstein, C.~Kohlf\"urst and M.~Zepf,
      [arXiv:1811.08895 [physics.optics]].
      
    \bibitem{Blinne:2018} 
      A.~Blinne, S.~Kuschel, S.~Tietze, and M.~Zepf,
      [arXiv:1801.0481 [physics.plasm-ph]].

      
    \bibitem{Gies:2017ezf}
      H.~Gies, F.~Karbstein, C.~Kohlf\"urst and N.~Seegert,
      Phys.\ Rev.\ D {\bf 97} (2018) no.7,  076002
      [arXiv:1712.06450 [hep-ph]].
      
    \bibitem{Gies:2017ygp}
      H.~Gies, F.~Karbstein and C.~Kohlf\"urst,
      Phys.\ Rev.\ D {\bf 97} (2018) no.3,  036022
      [arXiv:1712.03232 [hep-ph]].
      
    \bibitem{Heisenberg:1935qt} 
      W.~Heisenberg and H.~Euler,
      Z.\ Phys.\  {\bf 98}, 714 (1936), 
      an English translation is available at [physics/0605038].

    \bibitem{Schwinger:1951nm} 
      J.~S.~Schwinger,
      Phys.\ Rev.\ {\bf 82}, 664 (1951).
      
    \bibitem{Salamin:2006}
      Y.I. Salamin,
      Appl. Phys. B {\bf 86} (2007) no.2,  319--326.
      
    \bibitem{Waters:2017tgl} 
      W.~J.~Waters and B.~King,
      Laser Phys.\  {\bf 28}, 015003 (2018),
      [arXiv:1705.08554 [physics.optics]].
      
\bibitem{Tommasini:2010fb} 
  D.~Tommasini and H.~Michinel,
  Phys.\ Rev.\ A {\bf 82}, 011803 (2010)
  [arXiv:1003.5932 [hep-ph]].
  
\bibitem{King:2012aw} 
  B.~King and C.~H.~Keitel,
  New J.\ Phys.\  {\bf 14}, 103002 (2012)
  [arXiv:1202.3339 [hep-ph]].
  
\bibitem{King:2018wtn} 
  B.~King, H.~Hu and B.~Shen,
  Phys.\ Rev.\ A {\bf 98}, no. 2, 023817 (2018)
  [arXiv:1805.03688 [hep-ph]].

  \bibitem{Proceedings}
  A.~Blinne, H.~Gies, F.~Karbstein, C.~Kohlf\"urst and M.~Zepf,
  [arXiv:1812.02458 [hep-ph]].
  
\bibitem{Karbstein:2017jgh} 
  F.~Karbstein and E.~A.~Mosman,
  Phys.\ Rev.\ D {\bf 96}, no. 11, 116004 (2017)
  [arXiv:1711.06151 [hep-ph]].
  
    \bibitem{PhysRevLett.107.073602}
      Y. Monden and R. Kodama, 
      Phys. Rev. Lett. {\bf 107} (2011) 4,  073602.
      
    \bibitem{Richards358}
      B. Richards and E. Wolf.
      Proc. R. Soc. Lond. A (1959) 253 358-379.
  

\bibitem{KarbsteinSundqvist}
  F.~Karbstein and C.~Sundqvist,
  in preparation.
  
  \end{thebibliography}
\end{document}